\documentclass{emulateapj}
\usepackage{amsmath, natbib}
\begin{document}
\title{Systematic Effects in Extracting a {\em ``Gamma-Ray Haze''}  from Spatial Templates}
\author{Tim Linden$^{1}$ and Stefano Profumo$^{1,2}$}
\affil{$^1$ Department of Physics, University of California, Santa Cruz, 1156 High Street, Santa Cruz, CA, 95064}
\affil{$^2$ Santa Cruz Institute for Particle Physics, University of California, Santa Cruz, 1156 High Street, Santa Cruz, CA, 95064}
\slugcomment{Accepted by The Astrophysical Journal Letters}
\shortauthors{}

\begin{abstract}
Recent claims of a gamma-ray excess in the diffuse galactic emission detected by the Fermi Large Area Telescope made use of spatial templates from the interstellar medium (ISM) column density and the 408~Mhz sky as proxies for neutral pion and inverse Compton (IC) gamma-ray emission, respectively. We identify significant systematic effects in this procedure that can artificially induce an additional diffuse component with a morphology strikingly similar to the claimed gamma-ray haze. To quantitatively illustrate this point we calculate sky-maps of the ratio of the gamma-ray emission from neutral pions to the ISM column density, and of IC to synchrotron emission, using detailed galactic cosmic-ray models and simulations. In the region above and below the galactic center, the ISM template underestimates the gamma-ray emission due to neutral pion decay by approximately 20\%. Additionally, the synchrotron template tends to under-estimate the IC emission at low energies (few GeV) and to over-estimate it at higher energies (tens of GeV) by potentially large factors that depend crucially on the assumed magnetic field structure of the Galaxy. The size of the systematic effects we find are comparable to the size of the claimed ``Fermi haze'' signal. We thus conclude that a detailed model for the galactic diffuse emission is necessary in order to conclusively assess the presence of a gamma-ray excess possibly associated to the WMAP haze morphology.
\end{abstract}

\section{Introduction}
\label{sec:introduction}

One of the most exciting yet observationally challenging scientific objectives of the Large Area Telescope (LAT) on board the {\em Fermi Gamma-ray Space Telescope} \citep{Atwood:2009ez}, is the indirect detection of particle dark matter \citep{2008JCAP...07..013B}. However, limited gamma-ray statistics make diffuse signals arising from the pair-annihilation of dark matter difficult to differentiate from astrophysical processes. The limitation of using a diffuse signal to search for non-standard emission stems from difficulties in controlling the instrumental background and formulating a rigorous model for the astrophysical diffuse foregrounds. 

\begin{figure*}
		\plotone{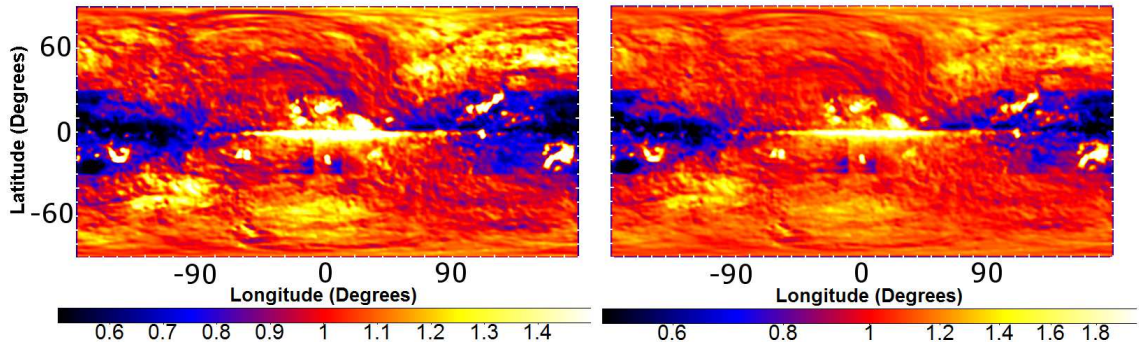}
		\caption{Normalized line-of-sight ratio of emission due to $\pi^0$ decay divided by the input line-of-sight gas density plotted on a linear scale (left) and logarithmic scale (right). Normalization is done on an equal area projection for $|$b$|$~$>$~5$^\circ$. }
		\label{pi0divgas}
\end{figure*}

An intriguing excess of microwave radiation in the WMAP data has been uncovered by \citet{Finkbeiner:2004us} and \citet{Dobler:2007wv}. The morphology and spectrum of the WMAP haze indicates a hard electron-positron injection spectrum spherically distributed around the galactic center. While the origin of this haze need not be related to {\em new} particle physics, the possibility that the WMAP haze corresponds to synchrotron radiation of stable leptons produced by dark matter has been explored in several studies \citep[see e.g.][]{Hooper:2007kb}. A potentially conclusive way to determine whether the WMAP haze originates from a population of energetic leptons is to observe gamma-rays produced by inverse Compton up-scattering (IC) of photons in the interstellar galactic radiation field (ISRF). 

Recently, \citet{Dobler:2009xz} (hereafter D09) examined the LAT gamma-ray sky and reported an excess emission morphologically similar to the WMAP haze. D09's observations suggest a confirmation of the {\em haze hypothesis}: that excess microwave emission stems from relativistic electron synchrotron with a spherical source distribution and a hard injection spectrum. In the ``Type 2" and ``Type 3" fits of D09, the excess was claimed over a best-fit background using spatial templates which employed the gas map of \citet{1998ApJ...500..525S} (SFD) to trace gamma-ray emission from $\pi^0$ decay, and the 408~Mhz Haslam synchrotron map \citep{1982A&AS...47....1H} to trace IC emission from galactic cosmic ray electrons. The spatial templates (plus an isotropic component obtained by mean-subtracting the residual skymap) were used to fit the observed gamma-ray sky in energy bins spanning 2-100~GeV. This analysis uncovered a residual gamma-ray emission above and below the galactic center with a morphology and spectrum similar to that found in the WMAP dataset \citep{2004ApJ...614..186F}.

In this $Letter$, we test the following assumptions used in D09 for the removal of astrophysical foregrounds at gamma-ray energies: 
\begin{itemize}
\item[(1)] that line of sight ISM maps are adequate tracers for the morphology of $\pi^0$ emission, and 
\item[(2)] that the 408 Mhz synchrotron map \citep{1982A&AS...47....1H} is an adequate tracer for the morphology of the galactic IC emission. 
\end{itemize}
Assumption (1) entails neglecting the morphology of galactic cosmic-ray sources, since the observed $\pi^0$ emission results from the line-of-sight integral of the gas density (``target'') times the cosmic-ray density (``beam''). Assumption (2) neglects the difference between the morphology of the ISRF and the galactic magnetic fields. 

On theoretical grounds, we expect that any detailed galactic cosmic-ray model would predict {\em systematic deviations} from the templates used in D09. Utilizing the galactic cosmic-ray propagation code {\tt Galprop}, we find that the procedure based on spatial templates creates deviations comparable to the amplitude of the D09 residual. Furthermore, we find that these deviations are morphologically similar to the Fermi haze. We thus conclude that the determination of an excess gamma-ray diffuse emission cannot reliably be assessed from the spatial template proxies used in the ``Type 2" and ``Type 3" fits of D09. We stress that our results do not claim that there is no ``haze'' in the Fermi data. In particular, the systematic effects we study here are not relavent to explain the puzzling excess emission in the ``Type 1'' fit of D09, which employes Fermi-LAT data in the 1-2 GeV range as a proxy for the morphology of the $\pi^0$ component. We comment on this ``Type 1'' approach in Section~\ref{sec:discussion}.

\begin{figure*}
		\plotone{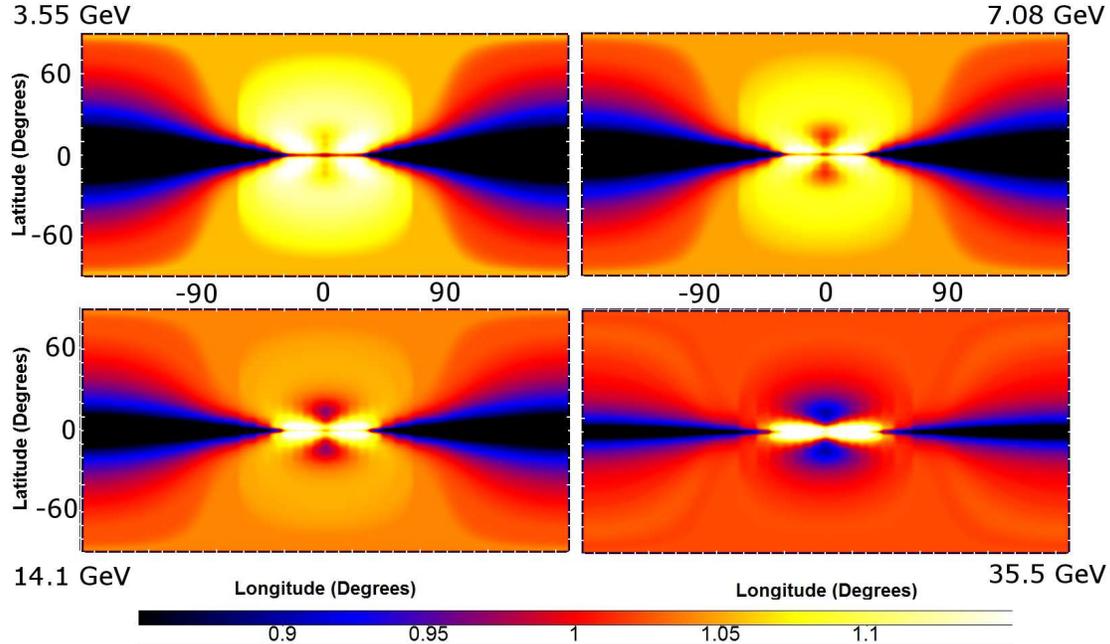}
		\caption{Normalized ratio of emission due to IC at four energy values (3.55~GeV (top left), 7.08~GeV (top right), 14.1~GeV (bottom left), 35.5~GeV (bottom right)) divided by synchrotron emission at 408 MHz. Normalization is done on an equal area projection for $|$b$|$~$>$~5$^\circ$.}
		\label{icsdivsync}
\end{figure*}

\section{Simulation Methods}

\label{sec:simulationmethods}
Employing the cosmic ray propagation code {\tt Galprop}~(v 50.1p) \citep{Strong:1998pw, 2009arXiv0907.0559S}, we compute the line-of-sight emission for galactic synchrotron, IC and $\pi^0$ decay predicted by a {\tt Galprop} model that is consistent with all cosmic ray and photon observations \citep[see][for further detail]{2009arXiv0907.0559S}. Except where noted, we employ standard parameters given by the GALDEF file 599278\footnote{http://galprop.stanford.edu/GALDEF/galdef\_50p\_599278} throughout this work.

A large uncertainty in the propagation of cosmic rays relates to the intensity and orientation of galactic magnetic fields \citep{1990ICRC....3..229B,1996ASPC...97..457H,1996A&A...308..433V} as the intensity of synchrotron radiation varies with the square of the local magnetic field intensity. In our default simulation we assume a magnetic field of random orientation and an intensity that exponentially decays in both $r$ and $z$ with scale radii of 10~kpc and 2~kpc respectively, normalized to 5~$\mu$G at the solar position \citep{Strong:1998pw, 2009arXiv0907.0559S}. 

To determine the accuracy of the D09 spatial templates for astrophysical foreground emission, we generate line-of-sight skymaps for the input gas density, as well as the outputs of emission due to $\pi^0$ decay, synchrotron and IC. Note that the gas density maps we employ here differ from the SFD map used in D09. Most notably, the SFD map traces dust, while our map traces galactic gas. The difference between these approaches is expected to be small, but might introduce additional systematic deviations. By dividing, pixel by pixel, the line-of-sight map for $\pi^0$ decay by the input gas map, and the map of IC emission by the synchrotron map, we can assess the size of any systematic effects produced by assumptions (1) and (2) of Section~\ref{sec:introduction}. We normalize each map over pixels of $|$b$|$~$>$~5$^\circ$, using equal area weighting to determine the normalization constant. This is equivalent to the masking procedure of D09 - though we do not mask out the galactic plane in our plots.

We select several regions of the sky for which we provide numerical analyses of each map we present, with a background normalized as in D09. We first evaluate D09's claim of an excess emission in the southern galactic plane between -30$^\circ$~$<$~b~$<$~-10$^\circ$ and $|$l$|$~$<$~15$^\circ$, indicated as the {\em D09 haze}. We add the symmetric {\em D09 northern haze} region (10$^\circ$~$<$~b~$<$~30$^\circ$ and $|$l$|<$15). D09 defines their haze to begin at 10$^\circ$, but only masked the region $|$b$|$~$<$~5$^\circ$. In order to determine the importance of this choice, we include both northern and southern regions following 5$^\circ$~$<$~$|$b$|$~$<$~25$^\circ$ and $|$l$|$~$<$~15$^\circ$, which we denote as the {\em Inner Haze} region. Since D09 models the morphology of the haze with a bivariate Gaussian that decays exponentially in both latitude and longitude, we further consider a map weighting the value at each unmasked pixel using a bivariate Gaussian of 25$^\circ$ in latitude and 15$^\circ$ in longitude, and dubbed the {\em Gaussian Haze}. Finally, in order to ascertain the variation of each map, we consider the galactic anticenter region 5$^\circ<|$b$|<$~25$^\circ$ and $|$l$|$~$>$~170. 

\section{Results}

\label{sec:results}
In Figure~\ref{pi0divgas} we show the normalized line-of-sight skymap for $\pi^0$ decay divided by the normalized line-of-sight gas density input into our {\tt Galprop} simulations. The results are shown on a linear scale (left) and a logarithmic scale (right), and both are smoothed using a Gaussian of 2$^\circ$ width. We find that the resultant skymap displays significant deviations from unity, with factors of approximately two above and below the galactic center, to values of about 0.3 near the galactic anti-center.

In Table~\ref{tab:regions} (top row), we provide both the average value in the D09 background region, as well as the numeric ratios in each of our defined regions. While this map for $\pi^0$ decay is taken at a test energy of 1~GeV, the variation of the ratio across four decades in energy (0.1 GeV - 1 TeV) is less than  2\%. While we find a deviation between the D09 haze region and the D09 background of only 4\%, we find a difference of 15\% between the southern inner haze region and the D09 background. These changes stem from the removal of a slight deficit in $\pi^0$ decay at approximately 30$^\circ$, and more importantly, the addition of increased emission between 5-10$^\circ$. Because the galactic gas distribution is not symmetric above and below the galactic plane, we expect some north-south variation, and we indeed find substantially higher values in the northern hemisphere. 

Finally, we calculate the ratio of the entire skymap weighted by the bivariate gaussian employed in D09. We find a 17\% excess in this measurement, which implies that using the gas map to account for the emission from $\pi^0$ decay in the Fermi-LAT signal would create a residual structure that could be fit by a bivariate gaussian with a mean intensity 17\% as large as the overall neutral pion gamma-ray emission. Since the latter vastly dominates the gamma-ray sky at low energies, this is a very significant effect between 0.1 and a few GeV. As remarked in \citet{jeanmarc}, the comparison of preliminary galactic gamma-ray diffuse models with the LAT source-subtracted sky data implies residuals which are likely much smaller than the systematic effect we point out here.

In Figure~\ref{icsdivsync} we show the normalized line-of-sight ratio of IC emission at four energies (3.55 GeV, 7.08 GeV, 14.1 GeV and 35.5 GeV) divided by synchrotron emission at 408 Mhz. The maps are again smoothed using a gaussian of 2$^\circ$, although in this case the intrinsic discreteness of {\tt Galprop} simulation grids makes the effect of this smoothing minimal. We note a pronounced structure above and below the galactic center which corresponds to an excess emission at low energies, but a deficit of emission at high energies.

\begin{figure*}
		\plotone{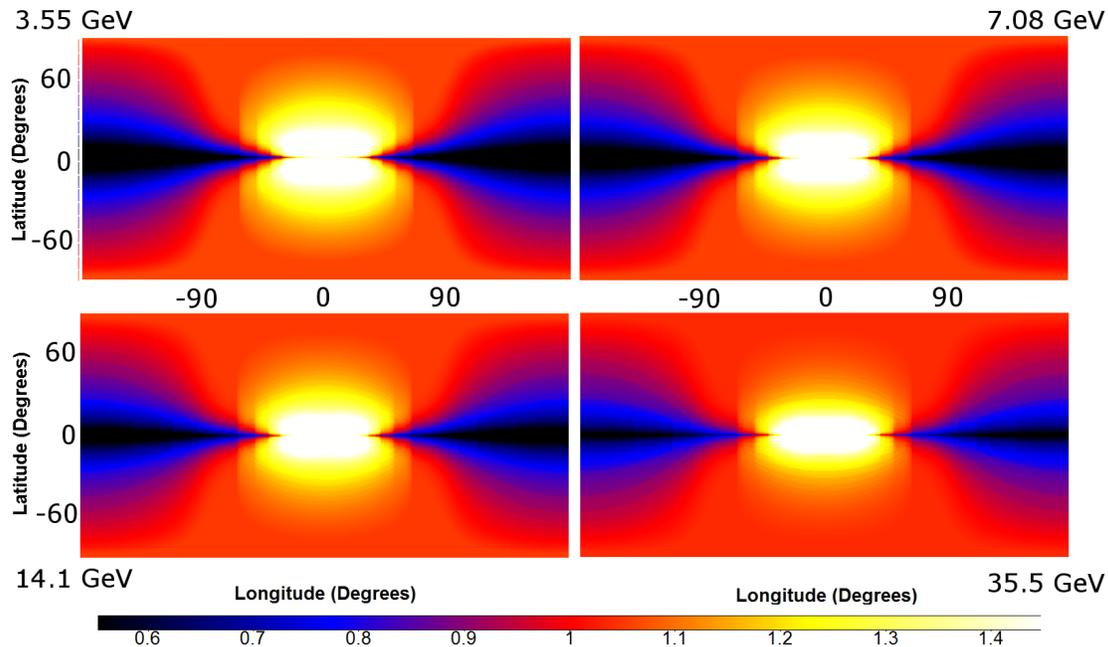}
		\caption{Same as Figure~\ref{icsdivsync} for a model with a magnetic field which decays exponentially in r and z with scale radii of 20~kpc and 2.0~kpc respectively and normalized to 5.0~$\mu$G at the solar position.}
		\label{icsdivsync.blargeradii}
\end{figure*}

Returning to Table~\ref{tab:regions} (Rows 2-5), we note that the ratios of IC and synchrotron emission remain relatively constant in all of the sky regions we consider. This occurs for two reasons: (1) the {\tt Galprop} model is nearly symmetric around the galactic plane, and (2) synchrotron emission dies off quickly at high latitudes, making the deviation just as prominant in the D09 regions as the inner regions. The morphology of these skymaps, however, has a pronounced energy dependence. At 3.55~GeV, there is a 16\% overabundance in IC emission above and below the galactic center, while at 35.5~GeV, this changes to an approximately 5\% deficit in IC emission. At all energies we see a deficit in IC near the galactic anti-center of between 5-17\%. The systematic effect stemming from assumption (2) would thus play a role in the determination of a gamma-ray haze spectrum, since the spatial template would change significantly with energy.

The morphology of these systematic effects is intuitively reasonable. Both IC and synchrotron emission depend on the same input population of high energy electrons (here modeled according to the results of \cite{fermiepem}). Thus, the morphology of the IC to synchrotron ratio depends primarily on the morphology of the energy density in the interstellar-radiation field (ISRF) compared to the galactic magnetic field (though we note that electrons at a given energy do not upscatter photons to a single energy, as the ISRF is composed of photons across a large range of frequencies). The magnetic field model we employ falls off sharply at high latitudes, but only weakly with changing radii, explaining its relative brightness near the galactic anti-center, and dimness at high galactic latitude. 

\section{Alternative Models}
\label{sec:parameterspace}

We note that the ratio of $\pi^0$ emission to the input gas map is fairly independent of the {\tt Galprop} propagation setup, as it depends primarily on the cosmic-ray density throughout space, which is constrained by local observations of cosmic ray fluxes and primary-to-secondary ratios. While the ratio of IC to synchrotron emission depends on the same source of high energy leptons, the morphology depends on the morphology of the ISRF compared to the magnetic fields.

Thus, one important feature which greatly affects the ratio of IC to synchrotron emission is the assumed morphology of the galactic magnetic fields, which are highly uncertain away from the Sun's local neighborhood \citep{1990ICRC....3..229B,1996ASPC...97..457H,1996A&A...308..433V}.  In Figure~\ref{icsdivsync.blargeradii}, we show the resulting ratio of IC to synchrotron emission for a magnetic field which decays exponentially with scale radii of 20.0~kpc in r and 2.0~kpc in z, tuned to a magnitude of 5~$\mu$G at the solar position. We note that this magnetic field setup increases the haze structure above and below the galactic pole by comparatively decreasing synchrotron radiation in that region, creating ratios greater than two in the galactic center and a gaussian deviation ranging from 1.54 at 3.55~GeV to 1.40 at 35.5~GeV. Furthermore, this setup decreases the height of the haze structure to approximately 45$^\circ$, which is in good agreement with the observed haze in D09. In Table~\ref{tab:regions} (Rows 6-9) we show the resulting haze ratios at each energy cutoff, finding ratios between 1.4-1.6 depending on the given energy level and region.

Another feature which may artificially yield greater uniformity in the ratio of the IC to synchrotron emission is the height of the diffusion zone. For small diffusion boxes, our high latitude gamma ray emission would be dominated by local cosmic rays, decreasing the variability of our results. However, this modeling trick cannot be used to support the Haze hypothesis, since the conjecture states that a flux of high energy leptons exists between 10$^\circ$ to 30$^\circ$ above and below the galactic center. A smaller diffusion region such as that at 2~kpc, will shut off emission at more than 14$^\circ$ latitude above the galactic plane, surpressing almost all emission from the haze region.

\section{Discussion} 
\label{sec:discussion}

\begin{figure*}
		\plotone{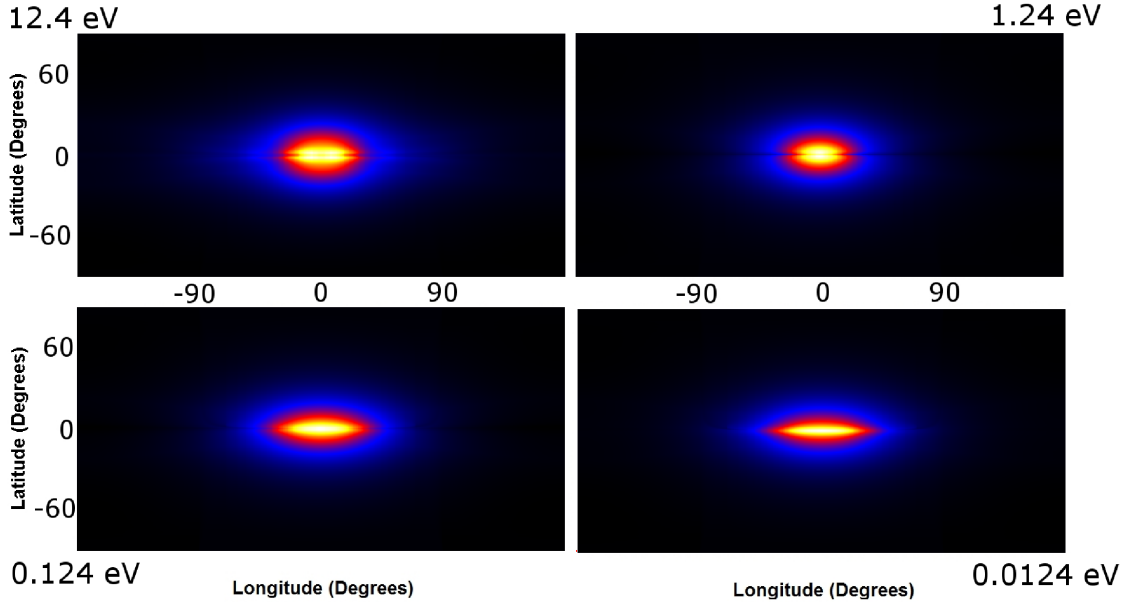}
		\caption{Intensity of the Interstellar Radiation function integrated across the line-of-sight, for energies of 12.4eV (top left), 1.24eV (top right), 0.124eV (bottom left), and 0.0124eV(bottom right). Plots are shown with arbitrary normalization.}
		\label{fig:isrf}
\end{figure*}
 
While a complete study of the diffuse emission detected by Fermi-LAT in terms of foreground templates is outside the scope of this $Letter$, work is currently ongoing that will also include the possible contribution to gamma radiation of nearby structures such as Loop I \citep{jeanmarc} which are not modeled within {\tt Galprop}. Whether improved spatial templates could conclusively pinpoint an excess diffuse emission in the Fermi-LAT data is hard to assess, as changes in the spatial morphology of each input will influence the best fit intensities of each component. However, we note that if the gas and synchrotron templates used in D09 were appropriate matches to $\pi^0$ and IC emission, we would expect ratios very close to unity across the entire skymap. We instead find significant deviations with an intensity comparable to the Fermi haze at low energies. Moreover, these deviations are not randomly distributed across the sky, but instead contain a structure resembling the bivariate gaussian reported by D09. Thus, our results point to the possibility that the Fermi haze determined by the ``Type 2'' and ``Type 3'' templates of D09 is the result of systematic effects in the spatial template fitting procedure as opposed to the existence of a new source class. 

We have shown that the use of the gas map would create a best match bivariate gaussian intensity approximately 17\% as strong as the $\pi^0$ decay amplitude. Since we expect $\pi^0$ decay to dominate the diffuse gamma-ray sky below 10~GeV, this error can explain a large fraction of the D09 haze. Similarly, IC due to ordinary galactic cosmic rays is expected to show an excess of approximately 20\% in the haze region, with large variances depending on the assumed magnetic field model. 

At high energies ($>$~10~GeV), $\pi^0$ decay is weaker, and thus a 15\% error is not expected to dominate astrophysical IC or haze signals. Furthermore, some magnetic field models create synchrotron templates which would overestimate IC emission in this region. Thus, the remaining amplitude of a D09 haze may in fact be larger or smaller than reported in this region. We note additional errors may be present at high energies both due to low photon counts as well as instrumental effects such as cosmic ray contamination. In addition, nearby cosmic rays such as those in the giant radio feature Loop I can fake a diffuse gamma-ray emission in the direction of the galactic center unrelated to a WMAP haze counterpart. 

D09 additionally adopts another template, dubbed ``Type I'', where they use the Fermi-LAT data in the 1-2 GeV range as a proxy for the morphology of $\pi^0$ emission. Using this template, D09 finds an excess emission that becomes more prominent at higher energies, indicating a residual with a harder spectrum than $\pi^0$-decay emission. They note that this excess has a morphology comparable with those found by their astrophysical template fits, except with a more peaked structure that may be due to the subtraction of the 1-2 GeV components of the ICS and bremsstrahlung maps. This emission template does not claim to subtract any $\gamma$-ray foregrounds except for those due to $\pi^0$-decay, but notes that the residual has a morphology which is peculiar for models such as ICS of high energy astrophysical leptons.

As this template does not make use of either the synchrotron or SFD maps, it falls outside the scope of our analysis. We agree that the emission is likely not due to $\pi^0$, and it is difficult to construct an astrophysical source distribution which is more pronounced towards high latitudes than near the galactic plane. Any galactic electron sources should lie close to the galactic plane, and would primarily upscatter starlight to GeV energies. In Figure~\ref{fig:isrf}, we show the ISRF used in our {\tt Galprop} models integrated over the line-of-sight. We see that the ISRF is also strongest along the galactic plane, extended in longitude rather than latitude by a ratio from 5-4 at 12.4eV to 9-1 at 0.0124eV. We further note that the ISRF dims by between 63-72\% between 10$^\circ$-30$^\circ$ latitude. This decay is itself slightly stronger than the decay in diffuse emission used in the D09 gaussian template.

This ISRF is morphologically identical to the IC morphologies obtained by comvolving the input ISRF with a isotropic and monochromatic input electron spectrum. Since this spectrum is much broader than the haze determined by D09, and falls slightly more quickly at high galactic latitudes, the input source class would have to be significantly peaked above and below the galactic plane, with very little extent along the plane. Furthermore, the source class must have a flux at 4.25~kpc above the galactic center, which is almost equvalent to the flux at 1.5~kpc above the galactic center, in order that the product of the lepton flux and ISRF fall off by a ratio comparable to the gaussian haze. The characteristics of this input source spectrum is not similar to those of expected galactic sources.

This problem night be overcome by several ad hoc changes in the ISRF around the galactic center region, or by changes in the convection currents and diffusion constants away from the galactic plane \citep{2009arXiv0910.2027G}. However, these same changes will greatly alter the assumed morphology of astrophysical ICS as well, possibly eliminating the need for an extra diffuse component. 

In summary, we showed that significant systematic effects make it difficult to reliably assess a diffuse gamma-ray emission in the region associated to the WMAP haze. A fully self-consistent galactic cosmic-ray model is necessary to model the astrophysical diffuse emission from the Galaxy and to compare it to the Fermi-LAT data.

\acknowledgements
We thank Troy Porter and Andy Strong for feedback regarding {\tt Galprop} models, as well as Gregory Dobler and Douglas Finkbeiner for useful discussions regarding the template construction in D09. TL is supported by a GAANN Fellowship by the Department of Education. SP is supported by an OJI Award from the US Department of Energy (DoE Contract DEFG02-04ER41268), and by NSF Grant PHY-0757911.

\newpage
\newpage
\begin{deluxetable}{ccccccccc}
\tablecolumns{8}
\tabletypesize{\footnotesize}
\tablewidth{0pt}
Model ID & Background &&D09 Haze &D09 N. Haze& Inner Haze& N. Inner Haze& Gaussian Haze& Anticenter\\
\\
\hline
\\
{\bf $\pi^0$ decay} &0.97 && 1.04 & 1.15 & 1.15 & 1.21 & 1.17 & 0.79 \\
\\
{\bf Base IC - 3.55 GeV}& 0.95 && 1.16 & 1.16 & 1.16 & 1.16 & 1.16 & 0.83\\
{\bf Base IC - 7.08 GeV}& 0.96 && 1.09 & 1.09 & 1.10 & 1.09 & 1.10 & 0.86\\
{\bf Base IC - 14.1 GeV}& 0.98 && 1.03 & 1.03 & 1.04 & 1.04 & 1.05 & 0.89\\
{\bf Base IC - 35.5 GeV}& 1.00 && 0.94 & 0.94 & 0.95 & 0.95 & 0.96 & 0.95\\
\\
{\bf Alt. IC - 3.55 GeV}& 0.88 && 1.47 & 1.46 & 1.57 & 1.56 & 1.54 & 0.62\\
{\bf Alt. IC - 7.08 GeV}& 0.89 && 1.42 & 1.41 & 1.53 & 1.51 & 1.49 & 0.65\\
{\bf Alt. IC - 14.1 GeV}& 0.90 && 1.38 & 1.37 & 1.49 & 1.48 & 1.46 & 0.68\\
{\bf Alt. IC - 35.5 GeV}& 0.92 && 1.31 & 1.30 & 1.43 & 1.41 & 1.40 & 0.74\\
\\
\hline
\\
\\
\caption{\label{tab:regions} Background (Column 1) and Ratios against Background (Columns 2-7) for regions specified by -30$^\circ<$b$<$-10$^\circ$ and $|$l$|<$15 (D09 Haze), 10$^\circ<$b$<$30$^\circ$ and $|$l$|<$15 (D09 North Haze), -25$^\circ<$b$<$-5$^\circ$ and $|$l$|<$15 (Inner Haze), 5$^\circ<$b$<$25$^\circ$ and $|$l$|<$15 (North Inner Haze),$|$b$|>$5 weighted by a bivariate gaussian of $\sigma_{b}$=25$^\circ$ and $\sigma_{l}$=15$^\circ$ (Gaussian Haze), and 5$^\circ<|$b$|<$25$^\circ$ and $|$l$|>$170 (Anticenter) for models of $\pi^0$ decay (Row 1), IC to synchrotron for a magnetic field with a intensity of 5$\mu$G at the solar position which exponentially decays in R and Z with scale heights of 10~kpc and 2~kpc respectively (Base, Rows 2-5), a magnetic field with a intensity of 5$\mu$G at the solar position which exponentially decays in R and Z with scale heights of 20~kpc and 2~kpc (Alt. IC, Rows 10-13)}
\end{deluxetable}




\end{document}